\documentclass[namedreferences,hyperref,optionalrh]{spr-sola}
\usepackage{graphicx}        
\usepackage{color}           
\usepackage{xcolor}   
\newcommand{\rev}[1]{{\color{black} #1}} 
\usepackage{lineno}

\newcommand{\kms} {km~$\rm s^{-1}$}

\usepackage{lineno}
\usepackage{soul}
\usepackage{verbatim}
\usepackage{textcomp}
\usepackage{wasysym}
\newcommand{\isis}{IS{\astrosun}IS}
\usepackage{float}
\usepackage[caption = false]{subfig}
\usepackage{paralist}

\chardef\us=`\_

\begin{document}

\begin{frontmatter}
\title{Observations of Switchback Chains in a Double Solar Proton Event}
\author[addressref=University of New Hampshire,corref,email={emily.mcdougall@unh.edu}]{\inits{E.O.}\fnm{Emily}~\snm{McDougall}\orcid{0009-0000-0011-6650}}
\author[addressref={University of New Hampshire},email={bala.poduval@unh.edu}]{\inits{B.}\fnm{Bala}~\snm{Poduval}\orcid{0000-0003-1258-0308}}

\address[id=University of New Hampshire]{Institute for the Study of Earth, Ocean, and Space Sciences, University of New Hampshire, 8 College Rd, Durham, NH, USA}

\runningauthor{Author-a et al.}
\runningtitle{\textit{Solar Physics} Example Article}

\begin{abstract}
Although recent research suggests a link in support of a model of switchback formation in the solar corona via interchange reconnection that is propagated outward with the solar wind and similarities in their ion composition to plasma instability produced plasmoids, these plasma instabilities have yet to be observationally linked to magnetic switchbacks. In this paper we aim to use the theoretical framework of a twin coronal mass ejection event which is known to include interchange reconnection processes and compare this model with experimental observations using Parker Solar Probe FIELDS and $\isis$ instrumentation
 of an actual event containing two CMEs identified via the associated solar proton event in order to further test and refine the hypothesis of interchange reconnection as a possible physical origin for the magnetic switchback phenomenon.  We also intend to introduce a plasma model for the formation of the switchbacks noted within the CME event.

\end{abstract}
\end{frontmatter}

\section{Introduction}
     \label{S-Introduction} 
The twin Coronal Mass Ejection (CME) phenomenon has been a well \rev{studied} phenomenon in the physics of the solar wind\rev{, with study over decades involving both observation and modelling of the phenomenon. While the twin CME phenomenon was first observed in coronagraphs made with the Large Angle and Spectrometric Coronagraph (LASCO) instrument \citep{Brueckner1995} on the Solar and Heliospheric Observatory (SOHO) \citep{Domingo1995}} and inferred from in situ measurements as far back as the 1970s \citep{Lugaz2017}, a more comprehensive physical model originates in \citep{Li2012}. This model predicts two CMEs erupting in sequence during a short interval of time from the same Active Region (AR) with a pseudo-streamer-like pre-eruption magnetic field configuration. The first CME in this scenario is \rev{typically} narrower and slower and the second CME, which is \rev{typically} wider and faster. 

\rev {While in an ordinary CME} the magnetic field configuration can lead to magnetic reconnection between the open and closed field lines that drape and enclose the first CME and its driven shock \citep{Owens2008}, \rev{in the twin CME model this is augmented by the presence of additional reconnection between the two CMEs}. The combined effect of the presence of the first shock and the existence of the open-close reconnection is that when the second CME erupts and drives a second shock, \rev{there is} both an excess of seed population and an enhanced turbulence level at the front of the second shock than the case of a single CME-driven shock, enhancing particle acceleration \citep{Dresing2022}. Further, work by \citep{Ding2014} notes that twin CME events with a lag time of less than 13 hours show a higher probability of leading to large \rev{Solar Energetic Particle (SEP)} events. \rev{SEPs typically result from either the site of explosive magnetic reconnection events known as solar flares or due to diffusive shock acceleration across shock waves associated with CMEs \citep{Reames2013}.}

Meanwhile, the magnetic switchback phenomenon has attracted a lot of attention from the scientific community since their initial observation in the Ulysses data \citep{Balogh1999,Neugebauer2013}, and subsequently in Helios 1 and 2 \citep{Borovsky2016,Horbury2018} and the Parker Solar Probe ($PSP$) \rev{\cite{Fox2016}}. $PSP$ observations were initially in November 2018 \citep{Bale2019, Kasper2019} and subsequently at other points as well \citep{Bowen2020,McComas2019,Schwadron2021,deWit2020,Horbury2020,Mozer2020,Rouillard2020,Tenerani2020}. These switchbacks \rev{are defined} as jumps in the plasma flow of the solar wind and a reversal of magnetic field orientation of at least $45^\circ$ from the background field, and a subsequent reversal to the background orientation over a period of greater than or equal to 10 seconds\rev{\citep{Fargette2022}}. Their precise physical origin remains poorly understood. Despite not being completely understood, they do have a few common characteristics including high Alv\'{e}nicity, an increase in the solar wind speed, and proton density. They typically contain ions at energy ranges detectable by $PSP$'s EPI-Lo instrument of the $\isis$ instrumentation suite\rev{, which detects ions in the energy range of 20 keV/nucleon to 15 MeV/nucleon, as opposed to the EPI-Hi instrument within the $\isis$ instrumentation suite which measures ions from 1 Mev/nucleon to 200 MeV/nucleon \citep{McComas2016,McComas2019}. The bulk of these ions in the switchback are protons.}

There are several competing hypotheses as to the physical origin of \rev{these} events. One such model originates from a study conducted by \rev{\cite{Neugebauer2012}} that notes that polar X ray jets were the source of velocity peaks associated with microstreams in the high speed solar wind. A model proposed by \rev{\cite{SterlingandMoore2015}} links such jets as a potential source of magnetic switchbacks. This was further refined with work done by \rev{\cite{Neugebauer2021}} noting a high correlation with microstreams and switchbacks that could potentially be explained by flux rope eruptions. 

Beyond that, switchbacks often exhibit both magnetic pressure \rev{($P_{mag}$)} and  magnetic normal and tangential components ($B_{N}$,$B_{T}$) variation as linked to the presence of a transverse bulk flow of around 20~\kms \rev{} in the heliospheric \rev{tangential} direction near perihelion \citep{Kasper2019}. Considering \rev{$P_{mag}$} variation is the defining signature of fast magnetosonic modes e.g., \citep{Lighthill1960}. It has been suggested on the basis of these observations that there is a general azimuthal (T direction) circulation of magnetic flux and plasma flow as a result of interchange reconnection in the low corona \citep{FiskandKasper2020}. This fits well with the noted increase in ion temperature, as this has been known to exist within magnetic reconnection events \citep{Gosling2007,Drake2009}. In addition, during such reconnection events hot tenuous loop plasma will be released, most likely in the form of a jet propagating at the Alfv\'{e}n speed relative to the background solar wind flow and in a direction that may not necessarily be radial, as well as a perturbed magnetic field in the transverse and radial fields with hot tenuous gas different from the surrounding ambient plasma \citep{Zank2020}. 

The ions that propagate with magnetic switchback events in such jets typically do not reverse direction with the change in direction of the magnetic field line \citep{Bandyo2021} due to the proton gyroradii in switchback events being much larger than the radius of curvature of the magnetic field lines. This is consistent with prior literature on the interchange reconnection model \citep{Fisk2005,FiskandSchwadron2001} and rules out a simple magnetic mirroring phenomenon \citep{Chen1984}. \rev{This phenomenon also is indicative of plasma that originates in the solar corona \citep{Phan2020}.} 

Further, ions in such events have been shown to \rev{be detected} at the Parker Solar Probe at \rev{points} that vary depending on the mass of the ion species \rev{even when they are at similar energies }\citep{McDougall2024}, indicating acceleration at a remote event rather than continuous acceleration, pointing against models such as interplanetary current sheet crossing, Alfv\'{e}nic turbulence, or shear forces between the fast and slow solar wind, which would not predict such behavior \citep{Squire2020,Schwadron2021}. At the same time however, ions do not appear to overtake the magnetic curvature of the switchback itself, and appear to propagate with it, suggesting some kind of \rev{ alternative plasma structure could potentially be trapping them.}  

In addition, switchbacks with a high energetic alpha particle concentration have been observed to be in the vicinity of CMEs \citep{McDougall2024}, suggesting a reconnection process is involved in their formation \citep{Aschwanden2019,Owens2011}. 

CMEs do possess flux rope structures that are very similar but of a different \rev{spatial} scale to plasmoids in the Earth's magnetotail\rev{ }\citep{Linton2009} that have been shown in Magnetohydrodynamic (MHD) and analytical models to exhibit dispersionless flux enhancements at multiple energies due to localized magnetic dips because of ion injections that can act as a kind of "pitch angle filter" for the ions contained in the plasmoids \citep{Gabrielse2023}. This could provide a mechanism for the observed behavior of the ions travelling with switchbacks, instead of traditional magnetic mirroring, when combined with magnetic tension due to the sharp changes in magnetic field direction. 

While the existing research so far points in the direction of magnetic switchbacks being a product of interchange reconnection, the physics of magnetic switchbacks in the context of the production of coronal mass ejection events has yet to  be studied, which is important in light of the discovery of switchbacks with alpha particles \rev{with energy levels significantly higher than the surrounding solar wind plasma} largely being found within several hours before or after CMEs. \rev{Due to the fact that as noted in \cite{Bandyo2021}, ions within the energy range investigated by \rev{EPI-Lo} are typical of magnetic switchbacks, an increase from this energy range into the range typically seen by \rev{EPI-Hi} could be indicative of the enhanced particle acceleration typically associated with a large SEP event generated by a twin CME}.

In this paper we attempt to use the well documented model of the twin CME to study the physics of the magnetic switchback in context, and to analyze whether the behavior observed in switchbacks does indeed conform to the predicted behavior of magnetic reconnection in the twin CME model. In doing so, we aim to gain a more nuanced understanding of magnetic switchbacks in the context of the physics of a twin CME. 

The paper is organized as follows: Section~\ref{sec:datamethod} presents the data and methodology we adopted for the current work. Our results are presented in Section~\ref{sec:results}, followed by a detailed discussion of the results and their theoretical implications in Section~\ref{sec:discussion}. Section~\ref{sec:conclusion} presents our conclusions and potential future work.

\section{Data/Methodology} \label{sec:datamethod}
\subsection{Data} \label{sub:data}

We utilize a single twin CME event using $PSP$ data that took place on August 18 2022 identified and plotted via the \rev{EPI-Hi} instrument data within the integrated Science Investigation of the Sun ($\isis$ instrument suite \rev{ \cite{McComas2016, McComas2019}}. We work primarily using The Energetic Particle Instrument high energy \rev{(EPI-Hi)} to identify energy levels higher than the low energy instrumentation \rev{(EPI-Lo)}  \url{https://spp-isois.sr.unh.edu/data_public/EPIHIlevel2/} that characterizes the energy range of magnetic switchbacks, as this can highlight energetic processes other than those which govern switchbacks, and isolate them as a separate physical process that accelerates the ions.  

We identify magnetic switchbacks using FIELDS \rev{\cite{Bale2016}} level 2 data from the Parker Solar Probe ($PSP$) within \rev{a radial(R), tangential (T), normal (N)} coordinate system from the Search Coil Magnetometer \rev{(SCM,\cite{Jannet2020})}
\url{http://research.ssl.berkeley.edu/data/psp/data/sci/fields/l2/mag_RTN/}. From there we use \rev{EPI-Lo} $\isis$ data to look at the plasma composition over time. The data is limited by the fact that large gaps in the $\isis$  data in general exist at irregular intervals, and as such the exact profile of the switchbacks within the event may be incomplete.

\rev{We also use Solar Probe Cup (SPC), or a Faraday cup instrument within the Solar Wind Electrons Alphas and Protons (SWEAP) instrument\rev{ }\citep{Case2020} to identify changes in the density of the plasma in order to determine underlying plasma structures in tandem with the magnetic field.}
\subsection{Method Adopted} \label{sub:method}

In the \rev{\cite{Li2012}} model of the twin CME scenario, we would see magnetic reconnection between the open and closed field lines of the first CME (preCME). The preCME would then provide both a seed population of pre-accelerated particles \rev{in its wake} with \rev{ a higher heavy element ratio that would be accelerated to higher energies}, and enhanced turbulence from the preceding shock.

This would require the Alfv\'{e}n wave turbulence to not decay away (estimated it takes \rev{9 hours} to do so) and consequently the twin-CME scenario requires a preCME with: \rev{a velocity greater than 300 \rev{km s-1} for shock generation}, a launch up to \rev{9 h} before the primary CME, and a centerline position angle (P.A.) within the span
of the primary CME \citep{Li2012, Kahler2014}. Further, we should see the front of the shock followed by a turbulent period and magnetic reconnection trailing the turbulence.
 
As such we use FIELDS data to identify turbulence in the event by noticing sharp rapid changes in the magnetic field lasting more than 3 seconds in duration. We then cross reference this event for recorded CME events  in the NASA Goddard Database of Notifications, Knowledge, Information (DONKI) database \url{https://kauai.ccmc.gsfc.nasa.gov/DONKI/search/} in order to verify the width, velocity, and launch time difference between the preCME and the primary CME to establish agreement with the \citep{Li2012} model.

Finally, we look at the magnetic switchback events surrounding the twin CME model to see if switchbacks correspond with \rev{ areas known to contain interchange reconnection in the model proposed by \cite{Li2012}, (Li model hereafter). In particular, the Li model predicts that the primary (second) CME overtaking the first (PreCME) can produce an overlapping shock, producing exceptionally large SEPs, potentially into the level of an SEP detectable by the earth's surface or a ground level enhancement (GLE). Such an overlapping shock would also allow us to see interchange reconnection events roughly equidistant from the overlapping shocks on the trailing edge of the flux rope of plasma following the turbulent area near the shock. Identifying switchbacks in this region could strengthen the argument that their origin lies in similar interchange reconnection events that occur in the solar corona.}  

\rev{Due to the variety of directions and angles represented in EPI-Lo data, all} \rev{EPI-Lo} particle flux data \rev{in Figure \ref{fig:Switchback}} is displayed on a symmetric log scale as to show deviations significant from the background radiation. 

\begin{figure}[htbp]
  \centering
  \centerline{\includegraphics[width=1.5\textwidth,clip=]{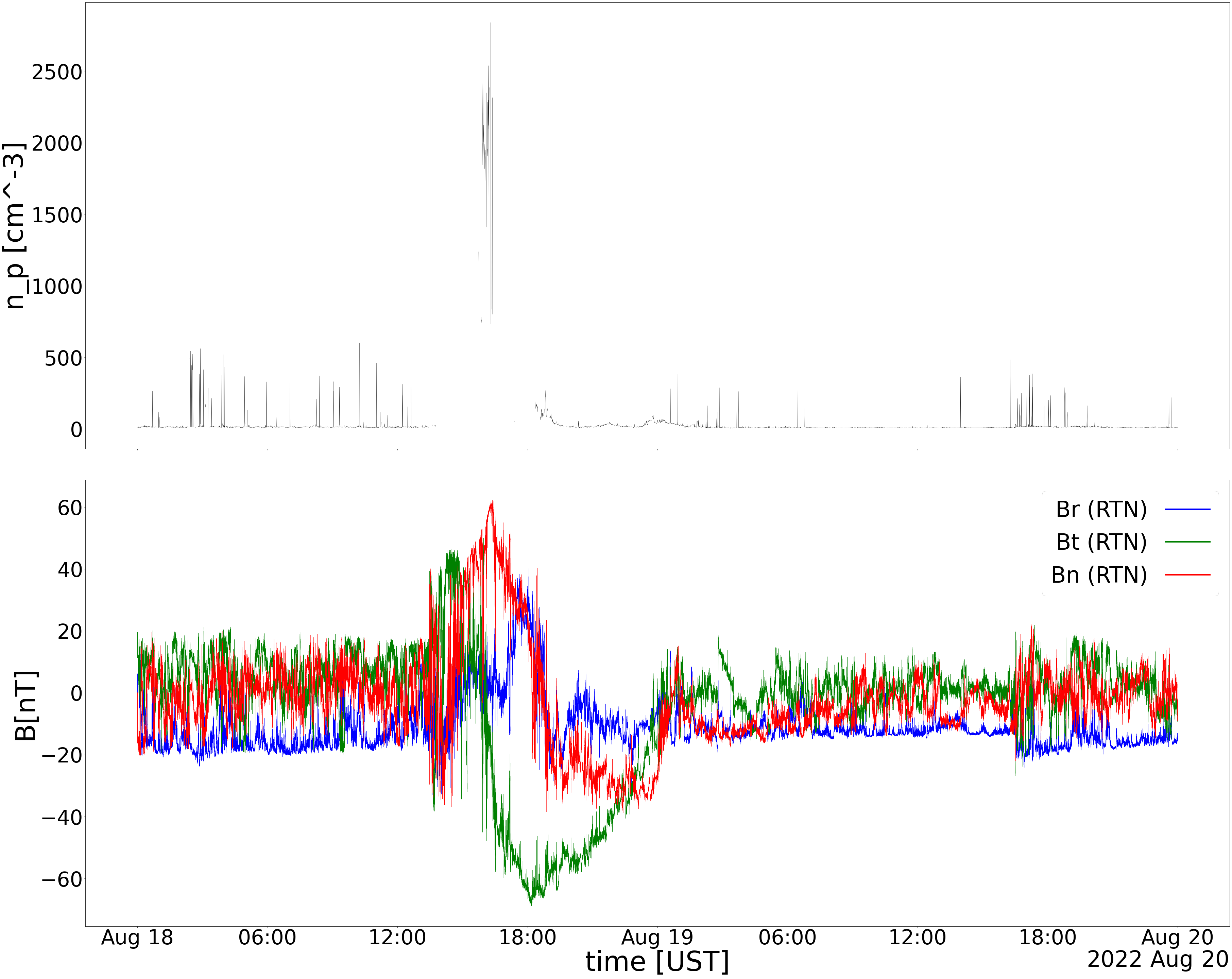}}
  \caption{ \rev{SPC and Fields data for the Twin CME event. The top plot depicts the proton density while the bottom plot depicts the magnetic field in RTN coordinates.} }
    \label{fig:SPC}
\end{figure}

\rev{The Fields and SPC data in Figure \ref{fig:SPC} are made using PYSPEDAS \citep{Angelopoulos2019}.}

\section{Results} \label{sec:results}

\begin{figure}[htbp]
  \centering
  \centerline{\includegraphics[width=1.1\textwidth,clip=]{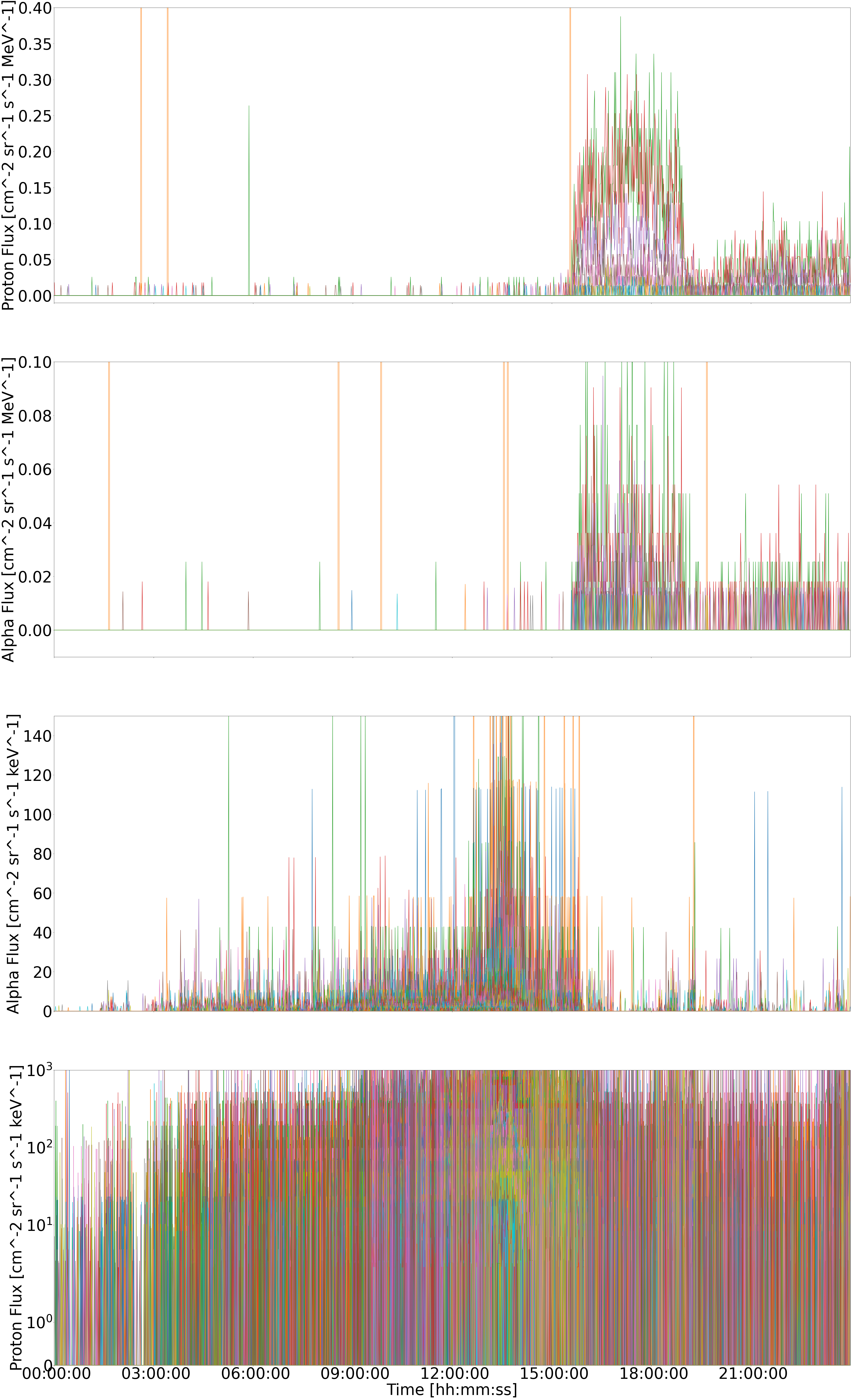}}
  \caption{ Parker Solar Probe \rev{$\isis$ data. First: EPI-Hi Proton Flux. Second: EPI-Hi Alpha Particle flux. Third: EPI-Lo Alpha Particle Flux. Fourth: EPI-Lo Proton Flux. Each color represents a different energy level from 1 MeV/nucleon to 200 MeV/nucleon for EPI-Hi and from 20 keV/nucleon to 15 MeV/nucleon for EPI-Lo.}
  }
    \label{fig:SPE}
\end{figure}

In cross referencing DONKI data, we find two CME events with both CMEs originating from active region 13708 with the preCME originating in the solar corona on \rev{2022 17 August}  at 15:24 \rev{UST} and the shock arriving at the $PSP$ on \rev{2022 18 August} at approximately \rev{5:00 UST}, with a turbulent sheath lasting around \rev{3 hours} until hitting a flux rope at approximately \rev{8:00 UST} \rev{corresponding with the low density cavity region of a CME thought to contain a flux rope} \citep{Kilpua2017}. These are depicted in \rev{Figure \ref{fig:SPC}}. The subsequent primary CME originated on \rev{2022 17 August} \rev{at} 14:53 \rev{UST} with the SEP event encompassing both CMEs 
depicted in \rev{Figure \ref{fig:SPE}}.

\begin{figure}[htbp]
\centering
  \centerline{\includegraphics[width=1.5\textwidth,clip=]{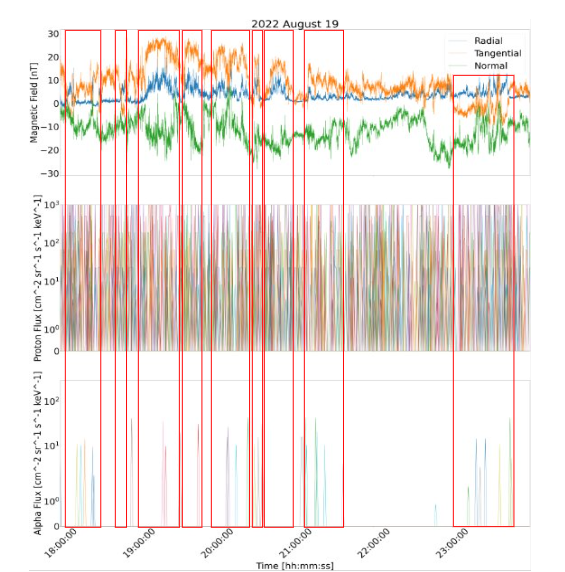}}
    \caption{ FIELDS and $\isis$ data for the Magnetic Switchbacks following the twin-CME event. \rev{Individual Switchbacks are marked within vertical red boxes. Each different color represents a different energy channel in the EPI-Lo instrument from 84.45 keV to 20.81 MeV for Alpha Particles and from 69.69 keV to 8 MeV for protons.} Top: Magnetic Field. Bottom Left: Proton Flux. Bottom Right: Alpha Particle Flux\rev{.} }
    \label{fig:Switchback}
\end{figure}

Further, we see multiple magnetic switchbacks surrounding the twin-CME event, but none of them seem to occur within the CME \rev{flux ropes and shocks} themselves, occurring before or after. The nearest switchback events  which were detectable with notable alpha flux in them were on \rev{2022 17} August from approximately 3:00 to 5:00 \rev{UST}, and several clustered switchbacks on \rev{2022 19} August shortly after 18:00 \rev{UST} depicted in \rev{Figure} \ref{fig:Switchback}, and on \rev{2022 20} August from 12:00 to 15:00 \rev{UST}. 

In each of our cases of switchback events found in the FIELDS data and the $\isis$ data the switchback event presents a proton flux far above the background solar wind levels\rev{.} The switchbacks also display a primary localization of the bulk of the proton flux at the initialization of the event, with the bulk of the alpha flux towards the end of the event, which is consistent with prior work \rev{by \cite{McDougall2024}} that suggests an interchange reconnection event in the solar corona that propagates outward in the solar wind as we see in Figure \ref{fig:Switchback}.

\section{Discussion} \label{sec:discussion}

The fact that the CMEs are more than 9 hours apart means that the Alfv\'{e}nic turbulence that characterized the \rev{\cite{Li2012}} model as a means of generating a seed population for the \rev{primary} CME as a means of generating a secondary SEP event likely doesn't apply for this particular event, and the higher energetic ion flux SEP appears to largely be a function of the larger \rev{primary} CME \rev{shown in Figure \ref{fig:SPE}}. 

\begin{figure}[htbp]
  \centering
  \centerline{\includegraphics[width=1.5\textwidth,clip=]{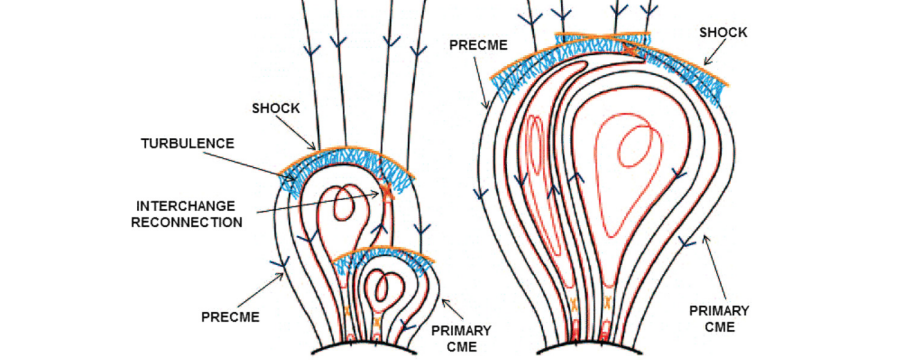}}
  \caption{ Model for a Twin CME Event adapted from \rev{\cite{Li2012}.}  }
    \label{fig:Twin_CME}
\end{figure}

The presence of a chain of switchbacks initiating around 18:00 \rev{UST} on \rev {2022 19 August} \rev{as depicted in Figure \ref{fig:Switchback}} is interesting because this corresponds with the period where the ion density begins a recovery from the dip in the density that follows the density spike near the shock, \rev{as we can see in Figure \ref{fig:SPC}} which is an area where we should expect magnetic interchange reconnection events after a CME as the closed field lines that make up the flux rope containing the plasma of the CME pinch off from the solar corona \rev{and reconnect with the open field lines of the solar wind}. Although this is analogous to the behavior in the \citep{Li2012} model, it is not precisely the exact same scenario due to the dissipation of Alfv\'{e}nic turbulence \rev{as noted in Figure} \ref{fig:Twin_CME}.

This \rev{scenario} is however, reminiscent of Particle In Cell (PIC) simulations on high \rev{Lundquist} number plasmas \rev{which are simulations involve computing the motion of individual particles in a Lagrangian frame by tracking them in continuous phase space, and quantities like densities and currents are computed simultaneously on stationary mesh points \citep{Dawson1983}. These simulations have} shown that high Lundquist number plasmas can reach an instability due to a super-Alfv\'{e}nic tearing instability \rev{where the high Alfv\'{e}n velocity and low diffusivity allows for magnetic pressure to dominate and generates reconnection along antiparallel field lines that form "magnetic islands" \citep{MacTaggart2020}. This increased reconnection rate allows} for a nonlinear reconnection rate larger than the \rev{reconnection rate of the standard Sweet-Parker model which assumes a simple 2D geometry with inflowing magnetic fields in the ±x direction and outward flow in the ±y direction \citep{Parker1957}. This increased reconnection rate} subsequently producing a rapid succession of plasmoids \citep{Bhatacharjee2009}. However, in order to obtain a tearing instability, a current sheet becomes unstable due the presence of both a finite conductivity and a reversed magnetic field \citep{Somov2006, MacTaggart2020}. This suggests that the reversed magnetic field formed as a result of interchange reconnection as the closed field lines around the CME reconnected with open field lines of the greater solar wind toward the rear of the CME \rev{as depicted in Figure }\ref{fig:Twin_CME} , changing the local magnetic topology. \rev{A similar phenomenon has been noted in magnetohydrodynamic numerical models that notes that in the wake of a CME production, the magnetic field lines in the solar corona become highly distorted. As the CME propagates outward, it leaves behind a region where the magnetic field lines reconnect and realign. This reconnection process generates a current sheet known as the Post CME Current sheet. Tearing instabilities can occur within this current sheet. As the magnetic field lines continue to evolve and reconnect, they can become unstable and develop into torn filaments. These torn filaments contribute to the broadening of the current sheet, and produce plasmoids within the current sheet\rev{ }\citep{Xie2022}. These switchback chains following the primary CME suggest that a similar process may have occurred, and the primary CME being outside the range to be interacting with the Alfv\'{e}nic turbulence left behind by the preCME suggests that this turbulence is more likely formed as part of the primary CME's formation and subsequent propagation outward from the corona, rather than a result of the post CME current sheet interacting with the turbulence left behind by the preCME.}

The clustering of switchbacks is also consistent with prior literature on the phenomenon \citep{deWit2020} and does bear resemblance to \rev{\cite{Zank2020}} where it is noted that the random walk of the magnetic field footpoints will sometimes bring the oppositely oriented open magnetic fields and loop magnetic fields sufficiently close to induce multiple successive temporally and spatially closely spaced interchange reconnection events. 

Moreover, the switchbacks in question also exhibit the same delay in arrival time at the $PSP$ between ion species that has been noted in prior switchbacks. The presence of such a delay in the switchback along with the lack of a tendency for the particle flux to move along the curves of the magnetic field suggests a remote origin for the acceleration of these ions that simply propagates outward with the solar wind, rather than something which originates via the motion of the solar wind itself, such as shear driven turbulence \citep{Squire2020,Schwadron2021}. The presence of higher energy alpha particle flux as well correlating with CMEs also tends to point towards structures produced in the corona rather than in the solar wind itself \citep{Owens2011}\rev{.}

Though, a simple magnetic mirroring mechanism to trap the ions so that they may propagate with the switchback is inconsistent with an ion population whose gyroradius is of a much different size than that would be expected for following along the curvature of the perturbed magnetic field.

A potential explanation involves plasmoids with diamagnetic motion induced dispersionless flux enhancements at multiple energies apart from an injection source of external ions. Due to particles with lower pitch angles not being trapped, the magnetic dips can act as a kind of pitch angle filter \citep{Gabrielse2023}.\rev{ }Such plasmoids have been shown to occur in what is known as a kinetic ballooning/interchange instability (BICI), which is a special case of a ballooning instability arising from internal plasma pressure from a curved magnetic field where the ballooning mode does not perturb the equilibrium magnetic field \citep{Hameiri1991}. 

BICIs have been shown in 3-D PIC simulations that a plasma sheet equilibrium with a minimum in the magnetic normal \rev{(Bz, \cite{Pritchett2013})} with a plasma flow out of the reconnection layer facilitating the removal of plasma and fast reconnnection\rev{ }\citep{Lyatsky2013}. As such, it remains a feasible mechanism for a reduction in the magnetic normal during magnetic erosion as the CME propagates through the interplanetary magnetic field \citep{Stamkos2023} to allow for the onset of interchange reconnection in the tail of the CME facing towards the sun, which produces a succession of plasmoids as the current sheets become unstable to wavenumber perturbations and produce a chain of plasmoids \citep{Loureiro2007} or more precisely, a tearing instability as the change in magnetic topology introduces a "pinching" effect, causing the tearing instability to produce series of plasmoids in a short amount of time \citep{Bhatacharjee2009}.

In addition, the behavior of the ion species as much higher than surrounding plasma is consistent with the model as \rev{\cite{Zank2020}} notes that during such events  hot tenuous loop plasma will be released, most likely in the form of a jet propagating at the Alfv\'{e}n speed relative to the background solar wind flow and in a direction that is not necessarily radial, as well as a perturbed magnetic field in the transverse and radial fields with hotter gas than the surrounding ambient plasma. 

It has also been noted previously in the literature that it is not clear that a linear magnetic reconnection event can propagate across the long distances required to be detectable by $PSP$. \rev{\cite{Drake2021}} proposes a model by which magnetic flux ropes created during interchange reconnection could provide for such long distance propagation. 

The larger scale magnetic structure that exists in the plasmoid chain near the flux ropes that make up the CMEs would appear to point more towards this \rev{flux rope} model as opposed to the simpler \rev{\cite{Zank2020} model}\rev{, particularly because the total magnetic field is constant as seen in Figure \ref{fig:Switchback}, a trait that is \rev{typical} of flux tubes \citep{Schüssler1993}. Moreover, their similarity to plasmoids can potentially be explained by the property of plasmoids to often align along the axes of flux ropes, effectively acting as a cross section of the flux rope \citep{Shibata2016}.} This is particularly relevant in light of literature that notes that coronal mass ejections and their interplanetary counterparts often show evidence of a twisted flux rope structure that is nearly identical, though of vastly different spatial scale, to plasmoids observed in the Earth’s magnetotail \citep{Linton2009}, which are also associated with BICIs. 

\rev{In addition}, the existence of a rapid succession of switchbacks that do not meet the criteria of traditional magnetic mirroring due to their gyroradii \citep{Bandyo2021} but do possess the constant total magnetic flux typical of a flux tube, suggests that these switchbacks may be similar to the plasmoids produced via BICI in the solar corona, \rev{which in turn may be} similar to the interchange instability produced plasmoids in the magnetotail that occur via interchange reconnection as a natural consequence of the solar wind pressure exerted on the day side magnetosphere moving towards the night side. 

Additional research however, would be needed to confirm \rev{that the plasmoids in the magnetotail, plasmoids in the corona, and switchbacks are all the same process}.  

\section{Concluding Remarks}\label{sec:conclusion}

We looked at the high alpha flux switchbacks in the vicinity of the twin-CME event and looked at the available DONKI data to check agreement with the Li model of a twin-CME event. Although we found that it was not within the bounds of a Li twin-\rev{CME} event due to the dissipation of Alfv\'{e}nic turbulence, there are things of value we can take away from this study.

The behavior of the particle flux and magnetic field components in these magnetic switchback events we observe in the $\isis$ and FIELDS data are consistent with a magnetic interchange connection model originating in the corona and propagating outward in the solar wind without further interaction with the magnetic field curvature. This is reminiscent of the model proposed by \rev{\cite{Zank2020}}. The data however, does not conflict significantly with the model proposed by \rev{\cite{Drake2021}}, and the presence of a larger magnetic structure in the plasmoid chains following the second CME does bolster the Drake model of flux ropes allowing for long distance propagation out toward the $PSP$ orbit. Additionally,the fact that many of these switchbacks take place in clusters in the immediate wake of the recovery of the ion density to normal solar wind levels suggests that these are connected to the formation of the CME event, which is consistent with prior research \cite{Aschwanden2019}.  

The results do suggest that switchbacks may occur during the formation of a CME with the main mechanism as reduction in the magnetic normal during magnetic erosion as the CME propagates through the interplanetary magnetic field to allow for the onset of interchange reconnection in the tail of the CME facing towards the \rev{S}un, producing a succession of plasmoids via a tearing instability in a short amount of time via the instability outlined in \rev{\cite{Loureiro2007}}. 

Moreover, as previously noted, the high frequency of switchbacks surrounding the CME events noticed is also reminiscent of \rev{modeling} done by \rev{\cite{Zank2020} that notes that chains of switchbacks could occur due to the random movements of magnetic footpoints}, and our study does not directly contradict this. However, \rev{the fact that the chain of switchbacks occurs directly at the end of the flux rope interior of the primary CME where the Li model suggests interchange reconnection would occur in a twin CME suggests that this may not necessarily be a result of a random process but of a specific instability}. Additionally, the \rev{\cite{Zank2020}} model fails to account for the instability of the magnetic kink propagating out to $PSP$ distances, which the flux rope model \rev{\cite{Drake2021}} \rev{does. Furthermore}, it remains unclear whether the microstreams described in the model proposed by \rev{\cite{Neugebauer2021}} being produced by a flux rope model would also be consistent with the jets predicted under \rev{\cite{Zank2020}}. 

However, we would be remiss if we did not acknowledge the limitations of this study.\rev{Although} we do observe switchbacks coalescing around a larger flux rope structure in the CMEs, and a mechanism for their formation that suggests their identity as plasmoids linking them to flux rope structures that suggests that \rev{\cite{Drake2021}} is more in line with the physical reality,  it is unclear what the limitations are of this model at this time and requires future study. It also remains to be seen exactly how similar the plasmoids in the magnetotail produced by this process are to the switchbacks observed in the solar wind. 

\section{Acknowledgements}

\noindent This work is supported by the National Science Foundation grant 2026579 
awarded to BP.

\rev{Parker Solar Probe was designed, built, and is now operated by the Johns Hopkins Applied
Physics Laboratory as part of NASA's Living with a Star (LWS) program (contract
NNN06AA01C). Support from the LWS management and technical team has played a critical
role in the success of the Parker Solar Probe mission. We would like to acknowledge the FIELDS team including the PI Stuart Bale, at the University of California, Berkeley, the $\isis$ team including the PI David McComas at Princeton University, the SPC team including the PI Justin Kasper at the University of Michigan and the Smithsonian Astrophysical Observatory, and NASA's Space Physics Data Facility for making the data publicly available. We would also like to acknowledge NASA for providing the available funding for these projects.}

\bibliographystyle{spr-mp-sola}

\bibliography{references.bib}

\begin{thebibliography}{64}
\ifx\bisbn     \undefined \def\bisbn  #1{ISBN #1}\fi
\ifx\binits    \undefined \def\binits#1{#1}\fi
\ifx\bauthor   \undefined \def\bauthor#1{#1}\fi
\ifx\batitle   \undefined \def\batitle#1{#1}\fi
\ifx\bjtitle   \undefined \def\bjtitle#1{\textit{#1}}\fi
\ifx\bvolume   \undefined \def\bvolume#1{\textbf{#1}}\fi
\ifx\byear     \undefined \def\byear#1{#1}\fi
\ifx\bissue    \undefined \def\bissue#1{#1}\fi
\ifx\bfpage    \undefined \def\bfpage#1{#1}\fi
\ifx\blpage    \undefined \def\blpage #1{#1}\fi
\ifx\burl      \undefined \def\burl#1{#1}\fi
\ifx\href      \undefined \def\href#1#2{#2}\fi
\ifx\betal     \undefined \def\betal{et al.}\fi
\ifx\bctitle   \undefined \def\bctitle#1{#1}\fi
\ifx\beditor   \undefined \def\beditor#1{#1}\fi
\ifx\bbtitle   \undefined \def\bbtitle#1{\textit{#1}}\fi
\ifx\bedition  \undefined \def\bedition#1{#1}\fi
\ifx\bseriesno \undefined \def\bseriesno#1{\textbf{#1}}\fi
\ifx\blocation \undefined \def\blocation#1{#1}\fi
\ifx\bsertitle \undefined \def\bsertitle#1{\textit{#1}}\fi
\ifx\bsnm      \undefined \def\bsnm#1{#1}\fi
\ifx\bsuffix   \undefined \def\bsuffix#1{#1}\fi
\ifx\bparticle \undefined \def\bparticle#1{#1}\fi
\ifx\barticle  \undefined \def\barticle#1{}\fi
\ifx\binstitute  \undefined \def\binstitute#1{#1}\fi
\ifx\bpublisher  \undefined \def\bpublisher#1{#1}\fi
\ifx\doiurl    \undefined \def\doiurl#1{\href{#1}{DOI}}\fi
\makeatletter
\def\safeHref#1#2#3{\in@{http}{#2}\ifin@\href{#2}{#3}\else\href{#1#2}{#3}\fi}
\makeatother
\ifx\adsurl    \undefined
  \def\adsurl#1{\safeHref{https://ui.adsabs.harvard.edu/abs/}{#1}{ADS}}\fi
\ifx\arxivurl  \undefined
  \def\arxivurl#1{\safeHref{http://arxiv.org/abs/}{#1}{arXiv}}\fi
\ifx\botherref \undefined \def\botherref#1{}\fi
\ifx\url       \undefined \def\url#1{#1}\fi
\ifx\bchapter  \undefined \def\bchapter#1{}\fi
\ifx\bbook     \undefined \def\bbook#1{}\fi
\ifx\bcomment  \undefined \def\bcomment#1{#1}\fi
\ifx\oauthor   \undefined \def\oauthor#1{#1}\fi
\ifx\citeauthoryear \undefined\def \citeauthoryear#1{#1}\fi
\def\endbibitem {}
\ifx\bconflocation  \undefined \def\bconflocation#1{#1} \fi

\bibitem[\protect\citeauthoryear{Angelopoulos et~al.}{2019}]{Angelopoulos2019}
\begin{barticle}
\bauthor{\bsnm{Angelopoulos}, \binits{V.}},
\bauthor{\bsnm{Cruce}, \binits{P.}},
\bauthor{\bsnm{Drozdov}, \binits{A.}},
\bauthor{\bsnm{Grimes}, \binits{E.W.}},
\bauthor{\bsnm{Hatzigeorgiu}, \binits{N.}},
\bauthor{\bsnm{King}, \binits{D.A.}},
\bauthor{\bparticle{et} \bsnm{al.}}:
\byear{2019},.
\bjtitle{{Space Sci Rev }}
\bvolume{{ 215}},
\bfpage{n/a}.
\doiurl{https://doi.org/10.1007/s11214-018-0576-4}.
\end{barticle}
\endbibitem

\bibitem[\protect\citeauthoryear{{Aschwanden}}{2019}]{Aschwanden2019}
\begin{bbook}
\bauthor{\bsnm{{Aschwanden}}, \binits{M.J.}}:
\byear{2019},
\bbtitle{{New Millennium Solar Physics}},
\bpublisher{Springer},
\blocation{New York, New York}.
\bisbn{3030139565}.
\end{bbook}
\endbibitem

\bibitem[\protect\citeauthoryear{Bale et~al.}{2016}]{Bale2016}
\begin{barticle}
\bauthor{\bsnm{Bale}, \binits{S.D.}},
\bauthor{\bsnm{Goetz}, \binits{K.}},
\bauthor{\bsnm{Harvey}, \binits{P.R.}},
\bauthor{\bsnm{Turin}, \binits{P.}},
\bauthor{\bsnm{Bonnell}, \binits{J.W.}},
\bauthor{\bparticle{Dudock~de} \bsnm{Wit}, \binits{T.}},
\bauthor{\bparticle{et} \bsnm{al.}}:
\byear{2016},.
\bjtitle{{Space Science Rev.}}
\bvolume{{204}},
\bfpage{49 }.
\end{barticle}
\endbibitem

\bibitem[\protect\citeauthoryear{Bale et~al.}{2019}]{Bale2019}
\begin{barticle}
\bauthor{\bsnm{Bale}, \binits{S.D.}},
\bauthor{\bsnm{Badman}, \binits{S.T.}},
\bauthor{\bsnm{Bonnell}, \binits{J.W.}},
\bauthor{\bsnm{Bowen}, \binits{T.A.}},
\bauthor{\bsnm{Burgess}, \binits{D.}},
\bauthor{\bsnm{Case}, \binits{A.W.}},
\bauthor{\bparticle{et} \bsnm{al.}}:
\byear{2019},.
\bjtitle{{Nature}}
\bvolume{{576}},
\bfpage{237}.
\doiurl{https://doi.org/10.1038/s41586-019-1818-7}.
\end{barticle}
\endbibitem

\bibitem[\protect\citeauthoryear{Balogh et~al.}{1999}]{Balogh1999}
\begin{barticle}
\bauthor{\bsnm{Balogh}, \binits{A.}},
\bauthor{\bsnm{Forsyth}, \binits{R.J.}},
\bauthor{\bsnm{Lucek}, \binits{E.A.}},
\bauthor{\bsnm{Hobury}, \binits{T.S.}},
\bauthor{\bsnm{Smith}, \binits{E.J.}}:
\byear{1999},.
\bjtitle{{Geophysical Research Letters}}
\bvolume{{26}},
\bfpage{631}.
\doiurl{https://doi.org/10.1029/1999GL900061}.
\end{barticle}
\endbibitem

\bibitem[\protect\citeauthoryear{Bandyopadhyay et~al.}{2021}]{Bandyo2021}
\begin{barticle}
\bauthor{\bsnm{Bandyopadhyay}, \binits{R.}},
\bauthor{\bsnm{Matthaeus}, \binits{W.H.}},
\bauthor{\bsnm{McComas}, \binits{D.J.}},
\bauthor{\bsnm{Joyce}, \binits{C.J.}},
\bauthor{\bsnm{Szalay}, \binits{J.R.}},
\bauthor{\bsnm{Christian}, \binits{E.R.}},
\bauthor{\bparticle{et} \bsnm{al.}}:
\byear{2021},.
\bjtitle{{A\&A}}
\bvolume{{650}},
\bfpage{6}.
\doiurl{https://doi.org/10.1051/0004-6361/202039800}.
\end{barticle}
\endbibitem

\bibitem[\protect\citeauthoryear{Bhatacharjee, Huang, and
  Yang}{2009}]{Bhatacharjee2009}
\begin{barticle}
\bauthor{\bsnm{Bhatacharjee}, \binits{A.}},
\bauthor{\bsnm{Huang}, \binits{Y.}},
\bauthor{\bsnm{Yang}, \binits{H.}}:
\byear{2009},.
\bjtitle{{Physics of Plasmas}}
\bvolume{{16}},
\bfpage{112102}.
\doiurl{https://doi.org/10.1063/1.3264103}.
\end{barticle}
\endbibitem

\bibitem[\protect\citeauthoryear{Borovsky and Denton}{2016}]{Borovsky2016}
\begin{barticle}
\bauthor{\bsnm{Borovsky}, \binits{J.E.}},
\bauthor{\bsnm{Denton}, \binits{M.H.}}:
\byear{2016},.
\bjtitle{{JGR Space Physics}}
\bvolume{{121}},
\bfpage{6107}.
\doiurl{https://doi.org/10.1002/2016JA022863}.
\end{barticle}
\endbibitem

\bibitem[\protect\citeauthoryear{Bowen et~al.}{2020}]{Bowen2020}
\begin{barticle}
\bauthor{\bsnm{Bowen}, \binits{T.A.}},
\bauthor{\bsnm{Mallet}, \binits{A.}},
\bauthor{\bsnm{Huang}, \binits{J.}},
\bauthor{\bsnm{Klein}, \binits{K.G.}},
\bauthor{\bsnm{Malaspina}, \binits{D.M.}},
\bauthor{\bsnm{Stevens}, \binits{M.}},
\bauthor{\bparticle{et} \bsnm{al.}}:
\byear{2020},.
\bjtitle{{ApJs}}
\bvolume{{246}},
\bfpage{66}.
\doiurl{https://doi.org/10.3847/1538-4365/ab6c65}.
\end{barticle}
\endbibitem

\bibitem[\protect\citeauthoryear{{Brueckner} et~al.}{1995}]{Brueckner1995}
\begin{barticle}
\bauthor{\bsnm{{Brueckner}}, \binits{G.E.}},
\bauthor{\bsnm{{Howard}}, \binits{R.A.}},
\bauthor{\bsnm{{Koomen}}, \binits{M.J.}},
\bauthor{\bsnm{{Korendyke}}, \binits{C.M.}},
\bauthor{\bsnm{{Michels}}, \binits{D.J.}},
\bauthor{\bsnm{{Moses}}, \binits{J.D.}},
\bauthor{\bparticle{et} \bsnm{al.}}:
\byear{1995},
\batitle{{The Large Angle Spectroscopic Coronagraph (LASCO)}}.
\bjtitle{\solphys}
\bvolume{162},
\bfpage{357}.
\doiurl{https://doi.org/10.1007/BF00733434}.
\adsurl{1995SoPh..162..357B}.
\end{barticle}
\endbibitem

\bibitem[\protect\citeauthoryear{Case et~al.}{2020}]{Case2020}
\begin{barticle}
\bauthor{\bsnm{Case}, \binits{A.W.}},
\bauthor{\bsnm{Kasper}, \binits{J.C.}},
\bauthor{\bsnm{Stevens}, \binits{M.L.}},
\bauthor{\bsnm{Korreck}, \binits{K.E.}},
\bauthor{\bsnm{Paulson}, \binits{K.}},
\bauthor{\bsnm{Daigneau}, \binits{P.}},
\bauthor{\bparticle{et} \bsnm{al.}}:
\byear{2020},.
\bjtitle{{ApJS}}
\bvolume{{246}},
\bfpage{43}.
\doiurl{https://doi.org/10.3847/1538-4365/ab5a7b}.
\end{barticle}
\endbibitem

\bibitem[\protect\citeauthoryear{Chen}{1984}]{Chen1984}
\begin{bbook}
\bauthor{\bsnm{Chen}, \binits{F.}}:
\byear{1984},
\bbtitle{Introduction to Plasma Physics and Controlled Fusion},
\bsertitle{Vol 1},
\bpublisher{Plenum, New York},
\bfpage{30}.
\bisbn{978-0-306-41332-2}.
\end{bbook}
\endbibitem

\bibitem[\protect\citeauthoryear{Dawson}{1983}]{Dawson1983}
\begin{barticle}
\bauthor{\bsnm{Dawson}, \binits{J.M.}}:
\byear{1983},.
\bjtitle{{Rev. Mod. Phys.}}
\bvolume{{55}},
\bfpage{403}.
\doiurl{https://doi.org/10.1103/RevModPhys.55.403}.
\end{barticle}
\endbibitem

\bibitem[\protect\citeauthoryear{Ding et~al.}{2014}]{Ding2014}
\begin{barticle}
\bauthor{\bsnm{Ding}, \binits{L.-G.}},
\bauthor{\bsnm{Li}, \binits{G.}},
\bauthor{\bsnm{Dong}, \binits{L.-H.}},
\bauthor{\bsnm{Jiang}, \binits{Y.}},
\bauthor{\bsnm{Jian}, \binits{Y.}},
\bauthor{\bsnm{Gu}, \binits{B.}}:
\byear{2014},.
\bjtitle{{JGR Space Physics}}
\bvolume{{119}},
\bfpage{1463}.
\doiurl{https://doi.org/10.1002/2013JA019745}.
\end{barticle}
\endbibitem

\bibitem[\protect\citeauthoryear{Domingo, Fleck, and
  Poland}{1995}]{Domingo1995}
\begin{barticle}
\bauthor{\bsnm{Domingo}, \binits{V.}},
\bauthor{\bsnm{Fleck}, \binits{B.}},
\bauthor{\bsnm{Poland}, \binits{A.I.}}:
\byear{1995},.
\bjtitle{{Sol Phys}}
\bvolume{{162}},
\bfpage{1}.
\doiurl{https://doi.org/10.1007/BF00733425}.
\end{barticle}
\endbibitem

\bibitem[\protect\citeauthoryear{Drake et~al.}{2009}]{Drake2009}
\begin{barticle}
\bauthor{\bsnm{Drake}, \binits{J.F.}},
\bauthor{\bsnm{Swisdak}, \binits{M.}},
\bauthor{\bsnm{Phan}, \binits{T.D.}},
\bauthor{\bsnm{Cassak}, \binits{M.A.}},
\bauthor{\bsnm{Shay}, \binits{M.A.}},
\bauthor{\bsnm{Lepri}, \binits{S.T.}},
\bauthor{\bparticle{et} \bsnm{al.}}:
\byear{2009},.
\bjtitle{{JGR Space Physics}}
\bvolume{{114}}.
\doiurl{https://doi.org/10.1029/2008JA013701}.
\end{barticle}
\endbibitem

\bibitem[\protect\citeauthoryear{Drake et~al.}{2021}]{Drake2021}
\begin{barticle}
\bauthor{\bsnm{Drake}, \binits{J.F.}},
\bauthor{\bsnm{Agapitov}, \binits{O.}},
\bauthor{\bsnm{Swisdak}, \binits{M.}},
\bauthor{\bsnm{Badman}, \binits{S.T.}},
\bauthor{\bsnm{Bale}, \binits{S.D.}},
\bauthor{\bsnm{Hobury}, \binits{T.S.}},
\bauthor{\bparticle{et} \bsnm{al.}}:
\byear{2021},.
\bjtitle{{A\&A}}
\bvolume{{650}},
\bfpage{8}.
\doiurl{https://doi.org/10.1051/0004-6361/202039432}.
\end{barticle}
\endbibitem

\bibitem[\protect\citeauthoryear{Dresing et~al.}{2022}]{Dresing2022}
\begin{barticle}
\bauthor{\bsnm{Dresing}, \binits{N.}},
\bauthor{\bsnm{Koloumvakos}, \binits{A.}},
\bauthor{\bsnm{Vainio}, \binits{R.}},
\bauthor{\bsnm{Rouillard}, \binits{A.}}:
\byear{2022},.
\bjtitle{{ApJL}}
\bvolume{{925}},
\bfpage{L21}.
\doiurl{https://doi.org/10.3847/2041-8213/ac4ca7}.
\end{barticle}
\endbibitem

\bibitem[\protect\citeauthoryear{Dudock~de Wit et~al.}{2020}]{deWit2020}
\begin{barticle}
\bauthor{\bparticle{Dudock~de} \bsnm{Wit}, \binits{D.T.}},
\bauthor{\bsnm{Krasnoselskikh}, \binits{V.V.}},
\bauthor{\bsnm{Bale}, \binits{S.D.}},
\bauthor{\bsnm{Bonnell}, \binits{J.W.}},
\bauthor{\bsnm{Bowen}, \binits{T.A.}},
\bauthor{\bsnm{Chen}, \binits{C.K.}},
\bauthor{\bparticle{et} \bsnm{al.}}:
\byear{2020},.
\bjtitle{{ApJS}}
\bvolume{{246}},
\bfpage{39}.
\doiurl{https://doi.org/10.3847/1538-4365/ab5853}.
\end{barticle}
\endbibitem

\bibitem[\protect\citeauthoryear{Fargette et~al.}{2022}]{Fargette2022}
\begin{barticle}
\bauthor{\bsnm{Fargette}, \binits{N.}},
\bauthor{\bsnm{Lavraud}, \binits{B.}},
\bauthor{},
\bauthor{\bsnm{Rouillard}, \binits{A.P.}},
\bauthor{\bsnm{Réville}, \binits{V.}},
\bauthor{\bsnm{Bale}, \binits{S.D.}},
\bauthor{\bsnm{Kasper}, \binits{J.C.}}:
\byear{2022},.
\bjtitle{{A\&A}}
\bvolume{{663}},
\bfpage{A109}.
\doiurl{https://doi.org/10.1051/0004-6361/202243537}.
\end{barticle}
\endbibitem

\bibitem[\protect\citeauthoryear{Fisk}{2005}]{Fisk2005}
\begin{barticle}
\bauthor{\bsnm{Fisk}, \binits{L.A.}}:
\byear{2005},.
\bjtitle{{ApJ}}
\bvolume{{626}},
\bfpage{563}.
\doiurl{https://doi.org/10.1086/429957}.
\end{barticle}
\endbibitem

\bibitem[\protect\citeauthoryear{Fisk and Kasper}{2020}]{FiskandKasper2020}
\begin{barticle}
\bauthor{\bsnm{Fisk}, \binits{L.A.}},
\bauthor{\bsnm{Kasper}, \binits{J.C.}}:
\byear{2020},.
\bjtitle{{ApJ}}
\bvolume{{894}},
\bfpage{L4}.
\doiurl{https://doi.org/10.3847/2041-8213/ab8acd}.
\end{barticle}
\endbibitem

\bibitem[\protect\citeauthoryear{Fisk and
  Schwadron}{2001}]{FiskandSchwadron2001}
\begin{barticle}
\bauthor{\bsnm{Fisk}, \binits{L.A.}},
\bauthor{\bsnm{Schwadron}, \binits{N.A.}}:
\byear{2001},.
\bjtitle{{ApJ}}
\bvolume{{560}},
\bfpage{425}.
\doiurl{https://doi.org/10.1086/322503}.
\end{barticle}
\endbibitem

\bibitem[\protect\citeauthoryear{Fox et~al.}{2016}]{Fox2016}
\begin{barticle}
\bauthor{\bsnm{Fox}, \binits{N.J.}},
\bauthor{\bsnm{Velli}, \binits{M.C.}},
\bauthor{\bsnm{Bale}, \binits{S.D.}},
\bauthor{\bsnm{Decker}, \binits{R.}},
\bauthor{\bsnm{Driesman}, \binits{A.}},
\bauthor{\bsnm{Howard}, \binits{R.A.}},
\bauthor{\bparticle{et} \bsnm{al.}}:
\byear{2016},.
\bjtitle{{Space Sci. Rev.}}
\bvolume{{204}},
\bfpage{7}.
\doiurl{https://doi.org/10.1007/s11214-015-0211-6}.
\end{barticle}
\endbibitem

\bibitem[\protect\citeauthoryear{Gabrielse et~al.}{2023}]{Gabrielse2023}
\begin{barticle}
\bauthor{\bsnm{Gabrielse}, \binits{C.}},
\bauthor{\bsnm{Gkioulidou}, \binits{M.}},
\bauthor{\bsnm{Merkin}, \binits{S.}},
\bauthor{\bsnm{Malaspina}, \binits{D.}},
\bauthor{\bsnm{Turner}, \binits{D.L.}},
\bauthor{\bsnm{Chen}, \binits{M.W.}}:
\byear{2023},.
\bjtitle{{Frontiers in Astronomy and Space Sciences}}
\bvolume{{10}}.
\doiurl{https://doi.org/10.3389/fspas.2023.1151339}.
\end{barticle}
\endbibitem

\bibitem[\protect\citeauthoryear{Gosling}{2007}]{Gosling2007}
\begin{barticle}
\bauthor{\bsnm{Gosling}, \binits{J.T.}}:
\byear{2007},.
\bjtitle{{ApJ}}
\bvolume{{671}},
\bfpage{L73}.
\doiurl{https://doi.org/10.1086/524842}.
\end{barticle}
\endbibitem

\bibitem[\protect\citeauthoryear{Hameiri, Laurence, and
  Mond}{1991}]{Hameiri1991}
\begin{barticle}
\bauthor{\bsnm{Hameiri}, \binits{E.}},
\bauthor{\bsnm{Laurence}, \binits{P.}},
\bauthor{\bsnm{Mond}, \binits{M.}}:
\byear{1991},.
\bjtitle{{JGR Space Physics}}
\bvolume{{ 96}},
\bfpage{1513}.
\doiurl{https://doi.org/10.1029/90JA02100}.
\end{barticle}
\endbibitem

\bibitem[\protect\citeauthoryear{Horbury, Matteini, and
  Stansby}{2018}]{Horbury2018}
\begin{barticle}
\bauthor{\bsnm{Horbury}, \binits{T.S.}},
\bauthor{\bsnm{Matteini}, \binits{L.}},
\bauthor{\bsnm{Stansby}, \binits{D.}}:
\byear{2018},.
\bjtitle{{MNRAS}}
\bvolume{{478}},
\bfpage{1980}.
\doiurl{https://doi.org/10.1093/mnras/sty953}.
\end{barticle}
\endbibitem

\bibitem[\protect\citeauthoryear{Horbury et~al.}{2020}]{Horbury2020}
\begin{barticle}
\bauthor{\bsnm{Horbury}, \binits{T.S.}},
\bauthor{\bsnm{Woolley}, \binits{T.}},
\bauthor{\bsnm{Laker}, \binits{R.}},
\bauthor{\bsnm{Matteini}, \binits{L.}},
\bauthor{\bsnm{Eastwood}, \binits{J.}},
\bauthor{\bsnm{Bale}, \binits{S.D.}},
\bauthor{\bparticle{et} \bsnm{al.}}:
\byear{2020},.
\bjtitle{{ApJS}}
\bvolume{{246}},
\bfpage{45}.
\doiurl{https://doi.org/10.3847/1538-4365/ab5b15}.
\end{barticle}
\endbibitem

\bibitem[\protect\citeauthoryear{Jannet et~al.}{2020}]{Jannet2020}
\begin{barticle}
\bauthor{\bsnm{Jannet}, \binits{G.}},
\bauthor{\bparticle{Dudock~de} \bsnm{Wit}, \binits{D.T.}},
\bauthor{\bsnm{Krasnoselskikh}, \binits{V.}},
\bauthor{\bsnm{Kretzschmar}, \binits{M.}},
\bauthor{\bsnm{Fergeau}, \binits{P.}},
\bauthor{\bsnm{Bergerard-Timofeeva}, \binits{M.}},
\bauthor{\bparticle{et} \bsnm{al.}}:
\byear{2020},.
\bjtitle{{JGR Space Physics}}
\bvolume{{126}},
\bfpage{n/a}.
\doiurl{https://doi.org/10.1029/2020JA028543}.
\end{barticle}
\endbibitem

\bibitem[\protect\citeauthoryear{Kahler and Vourlidas}{2014}]{Kahler2014}
\begin{barticle}
\bauthor{\bsnm{Kahler}, \binits{S.W.}},
\bauthor{\bsnm{Vourlidas}, \binits{A.}}:
\byear{2014},.
\bjtitle{{ApJ}}
\bvolume{{784}},
\bfpage{47}.
\doiurl{https://doi.org/10.1088/0004-637X/784/1/47}.
\end{barticle}
\endbibitem

\bibitem[\protect\citeauthoryear{Kasper et~al.}{2019}]{Kasper2019}
\begin{barticle}
\bauthor{\bsnm{Kasper}, \binits{J.C.}},
\bauthor{\bsnm{Bale}, \binits{S.D.}},
\bauthor{\bsnm{Belcher}, \binits{J.W.}},
\bauthor{\bsnm{Berthomier}, \binits{M.}},
\bauthor{\bsnm{Case}, \binits{A.W.}},
\bauthor{\bsnm{Chandran}, \binits{B.D.G.}},
\bauthor{\bparticle{et} \bsnm{al.}}:
\byear{2019},.
\bjtitle{{Nature}}
\bvolume{{576}},
\bfpage{228}.
\doiurl{https://doi.org/10.1038/s41586-019-1813-z}.
\end{barticle}
\endbibitem

\bibitem[\protect\citeauthoryear{Kilpua, Koskinen, and
  Pulkkinen}{2017}]{Kilpua2017}
\begin{barticle}
\bauthor{\bsnm{Kilpua}, \binits{E.}},
\bauthor{\bsnm{Koskinen}, \binits{H.E.J.}},
\bauthor{\bsnm{Pulkkinen}, \binits{T.I.}}:
\byear{2017},.
\bjtitle{{Living Rev Sol Phys}}
\bvolume{{14}},
\bfpage{5}.
\doiurl{https://doi.org/10.1007/s41116-017-0009-6}.
\end{barticle}
\endbibitem

\bibitem[\protect\citeauthoryear{Li et~al.}{2012}]{Li2012}
\begin{barticle}
\bauthor{\bsnm{Li}, \binits{G.}},
\bauthor{\bsnm{Moore}, \binits{R.}},
\bauthor{\bsnm{Mewaldt}, \binits{R.A.}},
\bauthor{\bsnm{Zhao}, \binits{L.}},
\bauthor{\bsnm{Labrador}, \binits{A.W.}}:
\byear{2012},.
\bjtitle{{Space Sci. Rev.}}
\bvolume{{171}},
\bfpage{141}.
\doiurl{https://doi.org/10.1007/s11214-011-9823-7}.
\end{barticle}
\endbibitem

\bibitem[\protect\citeauthoryear{Lighthill}{1960}]{Lighthill1960}
\begin{barticle}
\bauthor{\bsnm{Lighthill}, \binits{M.J.}}:
\byear{1960},.
\bjtitle{{RSPTA}}
\bvolume{{252}},
\bfpage{397}.
\doiurl{https://doi.org/10.1098/rsta.1960.0010}.
\end{barticle}
\endbibitem

\bibitem[\protect\citeauthoryear{Linton and Moldwin}{2009}]{Linton2009}
\begin{barticle}
\bauthor{\bsnm{Linton}, \binits{M.G.}},
\bauthor{\bsnm{Moldwin}, \binits{M.B.}}:
\byear{2009},.
\bjtitle{{J. Geophys. Res.}}
\bvolume{{114}},
\bfpage{A00B09}.
\doiurl{https://doi.org/10.1029/2008JA013660}.
\end{barticle}
\endbibitem

\bibitem[\protect\citeauthoryear{Loureiro, Schkekochihin, and
  Cowley}{2007}]{Loureiro2007}
\begin{barticle}
\bauthor{\bsnm{Loureiro}, \binits{N.F.}},
\bauthor{\bsnm{Schkekochihin}, \binits{A.A.}},
\bauthor{\bsnm{Cowley}, \binits{S.C.}}:
\byear{2007},.
\bjtitle{{Phys. Plasmas}}
\bvolume{{ 14}},
\bfpage{100703}.
\doiurl{https://doi.org/10.1063/1.2783986}.
\end{barticle}
\endbibitem

\bibitem[\protect\citeauthoryear{Lugaz et~al.}{2017}]{Lugaz2017}
\begin{barticle}
\bauthor{\bsnm{Lugaz}, \binits{N.}},
\bauthor{\bsnm{Temmer}, \binits{M.}},
\bauthor{\bsnm{Wang}, \binits{Y.}},
\bauthor{\bsnm{Farrugia}, \binits{C.}}:
\byear{2017},.
\bjtitle{{Sol Phys}}
\bvolume{{292}},
\bfpage{64}.
\doiurl{https://doi.org/10.1007/s11207-017-1091-6}.
\end{barticle}
\endbibitem

\bibitem[\protect\citeauthoryear{Lyatsky and Goldstein}{2013}]{Lyatsky2013}
\begin{barticle}
\bauthor{\bsnm{Lyatsky}, \binits{W.}},
\bauthor{\bsnm{Goldstein}, \binits{M.L.}}:
\byear{2013},.
\bjtitle{{Nonlin. Processes Geophys}}
\bvolume{{20}},
\bfpage{365}.
\doiurl{https://doi.org/10.5194/npg-20-365-2013}.
\end{barticle}
\endbibitem

\bibitem[\protect\citeauthoryear{MacTaggart}{2020}]{MacTaggart2020}
\begin{bbook}
\bauthor{\bsnm{MacTaggart}, \binits{D.}}:
\byear{2020},
\bbtitle{Topics in Magnetohydrodynamic Topology, Reconnection and Stability
  Theory},
\bsertitle{Vol 591},
\bpublisher{Springer, Cham.}
\bisbn{978-3-030-16343-3}.
\end{bbook}
\endbibitem

\bibitem[\protect\citeauthoryear{McComas et~al.}{2016}]{McComas2016}
\begin{barticle}
\bauthor{\bsnm{McComas}, \binits{D.J.}},
\bauthor{\bsnm{Alexander}, \binits{N.}},
\bauthor{\bsnm{Angold}, \binits{N.}},
\bauthor{\bsnm{Bale}, \binits{S.D.}},
\bauthor{\bsnm{Beebe}, \binits{C.}},
\bauthor{\bsnm{Birdwell}, \binits{B.}},
\bauthor{\bparticle{et} \bsnm{al.}}:
\byear{2016},.
\bjtitle{{Space Sci. Rev.}}
\bvolume{{204}},
\bfpage{187}.
\doiurl{https://doi.org/10.1007/s11214-014-0059-1}.
\end{barticle}
\endbibitem

\bibitem[\protect\citeauthoryear{McComas et~al.}{2019}]{McComas2019}
\begin{barticle}
\bauthor{\bsnm{McComas}, \binits{D.J.}},
\bauthor{\bsnm{Christian}, \binits{E.}},
\bauthor{\bsnm{Cohen}, \binits{C.M.S.}},
\bauthor{\bsnm{Cummings}, \binits{A.C.}},
\bauthor{\bsnm{Davis}, \binits{A.J.}},
\bauthor{\bsnm{Desai}, \binits{M.I.}},
\bauthor{\bparticle{et} \bsnm{al.}}:
\byear{2019},.
\bjtitle{{Nature}}
\bvolume{{576}},
\bfpage{223}.
\doiurl{https://doi.org/10.1038/s41586-019-1811-1}.
\end{barticle}
\endbibitem

\bibitem[\protect\citeauthoryear{McDougall and Poduval}{2024}]{McDougall2024}
\begin{botherref}
\oauthor{\bsnm{McDougall}, \binits{E.O.}},
\oauthor{\bsnm{Poduval}, \binits{B.}}:
2024,.
\textit{{ApJ}}
\textbf{{In Review}}.
\end{botherref}
\endbibitem

\bibitem[\protect\citeauthoryear{Mozer et~al.}{2020}]{Mozer2020}
\begin{barticle}
\bauthor{\bsnm{Mozer}, \binits{F.S.}},
\bauthor{\bsnm{Agapitov}, \binits{O.V.}},
\bauthor{\bsnm{Bale}, \binits{S.D.}},
\bauthor{\bsnm{Bonnell}, \binits{J.W.}},
\bauthor{\bsnm{Case}, \binits{T.}},
\bauthor{\bsnm{Chaston}, \binits{C.C.}}:
\byear{2020},.
\bjtitle{{ApJS}}
\bvolume{{246}},
\bfpage{68}.
\doiurl{https://doi.org/10.3847/1538-4365/ab7196}.
\end{barticle}
\endbibitem

\bibitem[\protect\citeauthoryear{Neugebauer}{2012}]{Neugebauer2012}
\begin{barticle}
\bauthor{\bsnm{Neugebauer}, \binits{M.}}:
\byear{2012},.
\bjtitle{{ApJ}}
\bvolume{{750}},
\bfpage{50}.
\doiurl{https://doi.org/10.1088/0004-637X/750/1/50}.
\end{barticle}
\endbibitem

\bibitem[\protect\citeauthoryear{Neugebauer and
  Goldstein}{2013}]{Neugebauer2013}
\begin{bchapter}
\bauthor{\bsnm{Neugebauer}, \binits{M.}},
\bauthor{\bsnm{Goldstein}, \binits{B.E.}}:
\byear{2013},
\bctitle{Double-proton beams and magnetic switchbacks in the solar wind}.
In: \bbtitle{Proceedings of the Thirteenth International Solar Wind Conference.
  AIP Conference Proceedings}
\bseriesno{1539},
\bfpage{46}.
\doiurl{https://doi.org/10.1063/1.4810986}.
\end{bchapter}
\endbibitem

\bibitem[\protect\citeauthoryear{Neugebauer and
  Sterling}{2021}]{Neugebauer2021}
\begin{barticle}
\bauthor{\bsnm{Neugebauer}, \binits{M.}},
\bauthor{\bsnm{Sterling}, \binits{A.C.}}:
\byear{2021},.
\bjtitle{{ApJL}}
\bvolume{{920}},
\bfpage{L31}.
\doiurl{https://doi.org/10.3847/2041-8213/ac2945}.
\end{barticle}
\endbibitem

\bibitem[\protect\citeauthoryear{Owens, Wicks, and Hobury}{2011}]{Owens2011}
\begin{barticle}
\bauthor{\bsnm{Owens}, \binits{M.}},
\bauthor{\bsnm{Wicks}, \binits{R.T.}},
\bauthor{\bsnm{Hobury}, \binits{T.S.}}:
\byear{2011},.
\bjtitle{{Sol Phys}}
\bvolume{{269}},
\bfpage{411}.
\doiurl{https://doi.org/10.1007/s11207-010-9695-0}.
\end{barticle}
\endbibitem

\bibitem[\protect\citeauthoryear{Owens et~al.}{2008}]{Owens2008}
\begin{barticle}
\bauthor{\bsnm{Owens}, \binits{M.}},
\bauthor{\bsnm{Crooker}, \binits{N.U.}},
\bauthor{\bsnm{Schwadron}, \binits{N.A.}},
\bauthor{\bsnm{Hobury}, \binits{T.S.}},
\bauthor{\bsnm{Yashiro}, \binits{S.}},
\bauthor{\bsnm{Xie}, \binits{H.}},
\bauthor{\bparticle{et} \bsnm{al.}}:
\byear{2008},.
\bjtitle{{GRL}}
\bvolume{{35}},
\bfpage{L20108}.
\doiurl{https://doi.org/10.1029/2008GL035813}.
\end{barticle}
\endbibitem

\bibitem[\protect\citeauthoryear{Parker}{1957}]{Parker1957}
\begin{barticle}
\bauthor{\bsnm{Parker}, \binits{E.N.}}:
\byear{1957},.
\bjtitle{{JGR}}
\bvolume{{62}},
\bfpage{509}.
\doiurl{https://doi.org/10.1029/JZ062i004p00509}.
\end{barticle}
\endbibitem

\bibitem[\protect\citeauthoryear{Phan et~al.}{2020}]{Phan2020}
\begin{barticle}
\bauthor{\bsnm{Phan}, \binits{T.D.}},
\bauthor{\bsnm{Bale}, \binits{S.D.}},
\bauthor{\bsnm{Eastwood}, \binits{J.P.}},
\bauthor{\bsnm{Lavraud}, \binits{B.}},
\bauthor{\bsnm{Drake}, \binits{J.F.}},
\bauthor{\bsnm{Oieroset}, \binits{M.}},
\bauthor{\bparticle{et} \bsnm{al.}}:
\byear{2020},.
\bjtitle{{ApJS}}
\bvolume{{246}},
\bfpage{34}.
\doiurl{https://doi.org/10.3847/1538-4365/ab55ee}.
\end{barticle}
\endbibitem

\bibitem[\protect\citeauthoryear{Pritchett and Coroniti}{2013}]{Pritchett2013}
\begin{barticle}
\bauthor{\bsnm{Pritchett}, \binits{P.L.}},
\bauthor{\bsnm{Coroniti}, \binits{F.V.}}:
\byear{2013},.
\bjtitle{{JGR Space Physics}}
\bvolume{{118}},
\bfpage{146}.
\doiurl{https://doi.org/1.1029/2012JA018143}.
\end{barticle}
\endbibitem

\bibitem[\protect\citeauthoryear{Reames}{2013}]{Reames2013}
\begin{barticle}
\bauthor{\bsnm{Reames}, \binits{D.V.}}:
\byear{2013},.
\bjtitle{{Space Sci. Rev.}}
\bvolume{{175}},
\bfpage{53}.
\doiurl{https://doi.org/10.1007/s11214-013-9958-9}.
\end{barticle}
\endbibitem

\bibitem[\protect\citeauthoryear{Rouillard et~al.}{2020}]{Rouillard2020}
\begin{barticle}
\bauthor{\bsnm{Rouillard}, \binits{A.P.}},
\bauthor{\bsnm{Kouloumvakos}, \binits{A.}},
\bauthor{\bsnm{Vourlidas}, \binits{A.}},
\bauthor{\bsnm{Kasper}, \binits{J.C.}},
\bauthor{\bsnm{Bale}, \binits{S.D.}},
\bauthor{\bsnm{Raouafi}, \binits{N.}},
\bauthor{\bparticle{et} \bsnm{al.}}:
\byear{2020},.
\bjtitle{ApJS}
\bvolume{{246}},
\bfpage{37}.
\doiurl{https://doi.org/10.3847/1538-4365/ab579a}.
\end{barticle}
\endbibitem

\bibitem[\protect\citeauthoryear{Schwadron and McComas}{2021}]{Schwadron2021}
\begin{barticle}
\bauthor{\bsnm{Schwadron}, \binits{N.A.}},
\bauthor{\bsnm{McComas}, \binits{D.J.}}:
\byear{2021},.
\bjtitle{\textit{arXiv:210203696 [astro-ph, physics:physics]}}
\bvolume{http://arxiv.org/abs/2102.03696},
\bfpage{n/a}.
\doiurl{https://doi.org/10.48550/arXiv.2102.03696}.
\end{barticle}
\endbibitem

\bibitem[\protect\citeauthoryear{Schüssler}{1993}]{Schüssler1993}
\begin{barticle}
\bauthor{\bsnm{Schüssler}, \binits{M.}}:
\byear{1993},.
\bjtitle{{Symposium-IAU}}
\bvolume{157},
\bfpage{27}.
\doiurl{https://doi.org/10.1017/S0074180900173826}.
\end{barticle}
\endbibitem

\bibitem[\protect\citeauthoryear{Shibata and Takasao}{2016}]{Shibata2016}
\begin{bbook}
\bauthor{\bsnm{Shibata}, \binits{K.}},
\bauthor{\bsnm{Takasao}, \binits{S.}}:
\byear{2016},
\bbtitle{Magnetic Reconnection: Concepts and Applications},
\bpublisher{Springer},
\blocation{Cham, Switzerland}.
\bisbn{978-3-319-26430-1}.
\end{bbook}
\endbibitem

\bibitem[\protect\citeauthoryear{Somov}{2006}]{Somov2006}
\begin{bbook}
\bauthor{\bsnm{Somov}, \binits{B.}}:
\byear{2006},
\bbtitle{Plasma Astrophysics Part II},
\bsertitle{Vol 341},
\bpublisher{Springer, NY, New York},
\bfpage{656}.
\bisbn{978-0-387-68894-7}.
\end{bbook}
\endbibitem

\bibitem[\protect\citeauthoryear{Squire, Chandran, and
  Meyrand}{2020}]{Squire2020}
\begin{barticle}
\bauthor{\bsnm{Squire}, \binits{J.}},
\bauthor{\bsnm{Chandran}, \binits{B.D.G.}},
\bauthor{\bsnm{Meyrand}, \binits{R.}}:
\byear{2020},.
\bjtitle{{ApJL}}
\bvolume{{891}},
\bfpage{L2}.
\doiurl{https://doi.org/10.3847/2041-8213/ab74e1}.
\end{barticle}
\endbibitem

\bibitem[\protect\citeauthoryear{Stamkos et~al.}{2023}]{Stamkos2023}
\begin{barticle}
\bauthor{\bsnm{Stamkos}, \binits{S.}},
\bauthor{\bsnm{Patsourakos}, \binits{S.}},
\bauthor{\bsnm{Vourlidas}, \binits{A.}},
\bauthor{\bsnm{Daglis}, \binits{I.A.}}:
\byear{2023},.
\bjtitle{{Sol Phys}}
\bvolume{{298}},
\bfpage{88}.
\doiurl{https://doi.org/10.1007/s11207-023-02178-7}.
\end{barticle}
\endbibitem

\bibitem[\protect\citeauthoryear{Sterling and
  Moore}{2015}]{SterlingandMoore2015}
\begin{barticle}
\bauthor{\bsnm{Sterling}, \binits{A.C.}},
\bauthor{\bsnm{Moore}, \binits{R.}}:
\byear{2015},.
\bjtitle{{Nature}}
\bvolume{{523}},
\bfpage{434}.
\doiurl{https://doi.org/10.1038/nature14556}.
\end{barticle}
\endbibitem

\bibitem[\protect\citeauthoryear{Tenerani et~al.}{2020}]{Tenerani2020}
\begin{barticle}
\bauthor{\bsnm{Tenerani}, \binits{A.}},
\bauthor{\bsnm{Velli}, \binits{M.}},
\bauthor{\bsnm{Matteini}, \binits{L.}},
\bauthor{\bsnm{R\'eville}, \binits{V.}},
\bauthor{\bsnm{Shi}, \binits{C.}},
\bauthor{\bsnm{Bale}, \binits{S.D..}}:
\byear{2020},.
\bjtitle{{ApJS}}
\bvolume{{246}},
\bfpage{32}.
\doiurl{https://doi.org/10.3847/1538-4365/ab53e1}.
\end{barticle}
\endbibitem

\bibitem[\protect\citeauthoryear{Xie et~al.}{2022}]{Xie2022}
\begin{barticle}
\bauthor{\bsnm{Xie}, \binits{X.}},
\bauthor{\bsnm{Mei}, \binits{Z.}},
\bauthor{\bsnm{Shen}, \binits{C.}},
\bauthor{\bsnm{Cai}, \binits{Q.}},
\bauthor{\bsnm{Ye}, \binits{J.}},
\bauthor{\bsnm{Reeves}, \binits{K.K.}},
\bauthor{\bparticle{et} \bsnm{al.}}:
\byear{2022},.
\bjtitle{{MNRAS}}
\bvolume{{509}},
\bfpage{406}.
\doiurl{https://doi.org/10.1093/mnras/stab2954}.
\end{barticle}
\endbibitem

\bibitem[\protect\citeauthoryear{Zank et~al.}{2020}]{Zank2020}
\begin{barticle}
\bauthor{\bsnm{Zank}, \binits{G.P.}},
\bauthor{\bsnm{Nakanotani}, \binits{M.}},
\bauthor{\bsnm{Zhao}, \binits{L.-L.}},
\bauthor{\bsnm{Adhikari}, \binits{L.}},
\bauthor{\bsnm{Kasper}, \binits{J.C.}}:
\byear{2020},.
\bjtitle{{ApJ}}
\bvolume{{903}},
\bfpage{1}.
\doiurl{https://doi.org/10.3847/1538-4357/abb828}.
\end{barticle}
\endbibitem

\end{thebibliography}


\end{document}